\pdfoutput=1
\documentclass[final, times, authoryear, 5p, twocolumn,12pt]{elsarticle}
\usepackage{epsfig, epstopdf, graphicx, color}
\usepackage{enumitem}
\usepackage{url}
\usepackage{amssymb}
\usepackage{amsfonts}
\usepackage{amsbsy}
\usepackage{amsmath}
\usepackage{natbib}
\usepackage[authoryear,round]{natbib}

\newcommand{\dg}[1]{{\it #1}}

\newcommand{\databank}{databank}

\newcommand{\gluevar}{Manager}

\newcommand{\spokes}{SPOKES}
\newcommand{\spokeslong}{{SP}ectr{O}scopic {KE}n {S}imulation}

\journal{Astronomy and Computing}

\begin{document}
\begin{frontmatter}
\title{ \spokes :
     an End-to-End Simulation Facility \\ for Spectroscopic Cosmological Surveys}

%-----------------------------------------------------------------------
% AUTHORS
%-----------------------------------------------------------------------
\author[fnal]{B. Nord} % Brian D. Nord
\ead{nord@fnal.gov}
\author[eth]{A. Amara}  % Adam Amara
\author[eth]{A. R{\'e}fr{\'e}gier} % Alexandre Refregier
\author[eth]{La. Gamper} % Laurenz Gamper
\author[eth]{Lu. Gamper} % Lukas Gamper
\author[eth]{B. Hambrecht} % Ben Hambrecht
\author[eth]{C. Chang} %Chihway Chang
\author[and]{J. E. Forero-Romero} % Jaime E. Forero-Romero
\author[ieec]{S. Serrano} % Santiago Serrano
\author[stan,kav]{C. Cunha} % Carlos Cunha
\author[ucl]{O. Coles} % Oliver Coles
\author[eth]{A. Nicola} % Andrina Nicola
\author[kav]{M. Busha} % Michael Busha
\author[ieec]{A. Bauer} % Ann Bauer
\author[aao]{W. Saunders} % Will Saunders
\author[ieec]{S. Jouvel} % Stephanie Jouvel
\author[ucl]{D. Kirk} % Donnacha Kirk
\author[stan,slac,kav]{R. Wechsler}%Risa Wechsler

%-----------------------------------------------------------------------
%ADDRESSES
%-----------------------------------------------------------------------
\address[fnal]{Fermilab Center for Particle Astrophysics, Fermi National Accelerator Laboratory, Batavia, IL 60510-0500}
\address[eth]{ETH Zurich, Department of Physics, Wolfgang-Pauli-Strasse 27, 8093 Zurich, Switzerland}
\address[stan]{Department of Physics, Stanford University, Stanford, CA 94305}
\address[slac]{SLAC National Accelerator Laboratory, 2575 Sand Hill Rd., MS 29, Menlo Park, CA 94025}
\address[itpz]{Institute for Theoretical Physics, University of Zurich,  8057 Zurich, Switzerland}
\address[kav]{Kavli Institute for Particle Astrophysics and Cosmology, 452 Lomita Mall, Stanford University, Stanford, CA, 94305}
\address[and]{Departamento de F\'{i}sica, Universidad de los Andes, Cra. 1 No. 18A-10, Edificio Ip, Bogot\'a, Colombia}
\address[ieec]{Institut de Ci\`{e}ncies de l'Espai, IEEC-CSIC, Campus UAB, Facultat de Ci\`{e}ncies, Torre C5 par-2, Barcelona 08193, Spain}
\address[aao]{Australian Astronomical Observatory, PO Box 915 North Ryde NSW 1670, Australia}
\address[ucl]{Department of Physics \& Astronomy, University  College London, Gower Street, London, WC1E 6BT, UK.}

%-----------------------------------------------------------------------
% ABSTRACT
%-----------------------------------------------------------------------
\begin{abstract}
The nature of dark matter, dark energy and large-scale gravity pose some of the most pressing questions in cosmology
today.  These fundamental questions require highly precise
measurements, and a number of wide-field spectroscopic survey instruments are being designed
to meet this requirement. A key component in these experiments is the development of
a simulation tool to forecast science performance, define requirement flow-downs,
optimize implementation, demonstrate feasibility, and prepare for exploitation.
We present \spokes\ (\spokeslong), an end-to-end simulation facility
for spectroscopic cosmological surveys designed to address this
challenge.  \spokes\ is based on an integrated
infrastructure, modular function organization, coherent data handling
and fast data access. These key features allow
reproducibility of pipeline runs, enable
ease of use and provide flexibility to update functions within the
pipeline.  The cyclic nature of the pipeline offers the
possibility to make the science output an efficient measure for
design optimization and feasibility testing.
We present the architecture, first science, and
computational performance results of the simulation pipeline. The
framework is general, but for the benchmark tests, we use the Dark
Energy Spectrometer (DESpec),
 one of the early concepts for the upcoming project, the Dark Energy
 Spectroscopic Instrument (DESI).
We discuss how the \spokes\ framework enables a rigorous
process to optimize and exploit spectroscopic survey experiments in order to derive
high-precision cosmological measurements optimally.

\end{abstract}

\end{frontmatter}

\clearpage

%-----------------------------------------------------------------------
\section{Introduction}
%-----------------------------------------------------------------------
Progress in cosmology over recent decades has led to some of the most pressing questions in
fundamental science today, such as those related to the nature of dark matter, dark energy,
and gravity on cosmological scales. To address these questions, several wide-field
spectroscopic surveys are in progress or being planned, including WiggleZ \citep{
2010MNRAS.401.1429D}, the Hobby-Eberly Telescope Dark Energy EXperiment
\citep[HETDEX;][]{Adams:2010gk}, the Prime Focus Spectrograph
\citep[PFS;][]{2012arXiv1206.0737T}, the Big Baryon Oscillation Spectroscopic Survey
\citep[BigBOSS;][]{2011arXiv1106.1706S}, the Dark Energy Spectrometer
\citep[DESpec;][]{Abdalla:2012tm}, the Dark Energy Spectroscopic Instrument
(DESI\footnote{\tt http://desi.lbl.gov}) and the 4m Multi-Object Spectroscopic Telescope
\citep[4MOST;][]{2012SPIE.8446E..0TD}. The goal of these experiments is to provide
three-dimensional maps of the large-scale structure of the universe by measuring the angular
positions and redshifts of galaxies in large cosmological volumes.

To reach the levels of precision that are needed to address the fundamental open questions
in cosmology, these experiments must meet stringent requirements on both statistical power
and control of systematic errors. These requirements, therefore, drive all aspects of these
experiments from instrument design to survey optimization.  Simulation tools play a key role
in the design and optimization process: they are important for forecasting science
performance of a given experimental configuration and, moreover, to demonstrate the
feasibility of a mission design. Rigorous simulation tools can also allow the science team
to prepare for the science interpretation and exploitation of the data.

Such simulation tools have been developed for a number of cosmological surveys. For example,
the optical imaging project, Sloan Digital Sky Survey \citep[SDSS;][]{2000AJ....120.1579Y}
employed the Monte-Carlo technique to test deblending in the image processing pipeline prior
to the survey taking
place\footnote{\url{http://www.astro.princeton.edu/~rhl/photo-lite.pdf}}. SDSS also used
simulations of galaxies, stars and QSOs to prepare and calibrate analysis pipelines for
object classification \citep{1999AJ....117.2528F,2001AJ....122.1861S}, and for measurements
of the galaxy luminosity function \citep{2001AJ....121.2358B}.

The Dark Energy Survey \citep[DES;][]{2012SPIE.8446E..11F} will rely on large-scale
simulated catalogs to forecast cosmological constraints
\citep[e.g.,][]{2012ApJ...753..152B}, develop science analysis pipelines
\citep[e.g.,][]{2014arXiv1411.0032C}, and improve the survey strategy
\citep{2012ASPC..461..201N}. Galaxy catalogs, along with pixel-level image simulations, also
permit the development of image reduction pipelines.  Next-generation experiments, like the
Large Synoptic Survey Telescope \citep[LSST;][]{2012arXiv1211.0310L}, will employ
photon-level simulations to account for sources of noise, such as atmospheric turbulence
\citep{Connolly:2010fa, 2012SPIE.8444E..4PC,2012arXiv1206.1378C}.  In addition, operational
procedures, like survey strategy, have benefited from extensive simulations
\citep{2006SPIE.6270E..45D, 2009arXiv0912.0201L, 2011ASPC..442..329G, 2012SPIE.8451E..12H}.

As simulations and forward modeling methods play an increasingly important role in survey
design and analysis, new frameworks  for simulations have been developed. For example, the
Monte-Carlo-Control-Loop (MCCL) method, proposed by \cite{Refregier:2013un}, aims to build a
robust set of control loops, based on simulations, for verifying that complex measurement
methods meet systematic requirement levels. Such system-level optimizations have underscored
the need for fast simulations leading to efforts to develop simulations that are fast enough
to support such integrated development. An example of this is Ultra Fast Image Generator
\citep[UFig;][]{2013A&C.....1...23B}, which has been developed to quickly and efficiently
produce simulated wide-field survey images.

Spectroscopic surveys can take advantage of the same kinds of mock galaxy catalogs as
imaging surveys for forecasting. However, there are more operations and additional levels of
complexity in spectroscopic surveys: for example, targets must be pre-selected before the
surveying can begin, and for each tile on the sky, fibers are allocated to sources;
moreover, these operations are intertwined, such that decisions regarding one will affect
one or more of the others.

In response to these challenges, spectroscopic experiments have undertaken several design
approaches. For example, some recent surveys, such as SDSS-III's Baryon Oscillation
Spectroscopic Survey (BOSS) and the 6dF Galaxy Survey (6dFGS), focused simulation efforts
toward optimizing the fiber allocation and tiling algorithms \citep{Campbell:2004cha,
2003AJ....125.2276B}. Studies for BigBOSS have performed target selection on mock catalogs
and simulated two-dimensional images of galaxy spectra in an effort to develop the tools to
extract spectra from images \citep{2011arXiv1106.1706S}.  4MOST has developed the Facility
Simulator \citep{2012SPIE.8448E..0XB}, which links together the survey strategy and fiber
allocation to convert an input catalog (from an imaging survey) into one that would result
from a 4MOST survey.

In this paper, we describe the \spokeslong\ (\spokes), an end-to-end simulation facility for
spectroscopic cosmological surveys. \spokes\ is built on an integrated infrastructure,
modular function organization, coherent data handling, and fast data access. These key
features allow reproducibility of pipeline runs, enable ease of use, and provide flexibility
to update functions within the pipeline.  The pipeline's framework is also cyclic: it can be
easily executed in a loop, offering the possibility to make the science output an efficient
measure for design optimization and feasibility testing.  While the framework is general, we
use the design of the DESpec experiment concept \citep{Abdalla:2012tm} as a baseline for
development and for benchmarking results. DESpec was one of the early concepts for the
upcoming DESI experiment.

We present here the architecture, and the science and computational performance results of
the \spokes\ simulation pipeline. \spokes\ and all the modules are written in the Python
programming language\footnote{\url{http://www.python.org}}.

The paper is organized as follows. In \S\ref{sec:challenges}, we describe the challenges
that spectroscopic surveys need to meet in order to reach the required precision, as well as
the principal ingredients in a framework that simulates surveys. In \S\ref{sec:solution}, we
present \spokes\  and show how its design addresses these challenges.  In
\S\ref{sec:results}, we present science and performance results of the simulation pipeline.
Our conclusions are summarized in \S\ref{sec:summary}. Details regarding the data format
choices and the input cosmological simulation are described in the Appendix.

%-----------------------------------------------------------------------
\section{Challenges for Spectroscopic Survey Simulations}\label{sec:challenges}
% -----------------------------------------------------------------------

\subsection{Challenges} Future wide-field spectroscopic surveys offer great promise to
address the fundamental questions described above. Their exploitation, however, will pose
the following challenges that need to be addressed in order to achieve the required
accuracy.

\begin{itemize}

\item {\it High precision:}  The next generation of spectroscopic surveys,
along with other Stage IV dark energy experiments \citep{2006astro.ph..9591A}, aim to
measure the dark energy equation of state parameter, $w$, to percent-level precision. This
sets ambitious requirements---e.g., that these experiments cover large cosmological volumes
and maintain tight control over errors.

\item {\it Systematics:} As the statistical power of surveys increases, numerous sources of
systematic errors become significant. These include errors in the calibration of the survey
selection function, inhomogeneous photometric target selection, masking, etc. These
systematics need to be carefully calibrated and controlled so that they become subdominant
compared to statistical errors.

\item {\it Complexity:} The difficulty in controlling systematic errors is compounded by the
fact that errors can couple to one another in a non-linear fashion in spectroscopic surveys.
For example, there is an an interplay between target selection and fiber allocation, as each
type of target (e.g., luminous red galaxy, emission line galaxy or QSO) will generally
require observation through a different wavelength range.  Each fiber is attached to a
spectrograph with a specific wavelength range.  Therefore, there must be enough fibers of
each type (i.e., of each wavelength range) available for each type of target selected.
Unless such effects and couplings are well modeled and tested, there is a risk that these
effects will be imprinted on the final measured galaxy correlation function. This would then
lead to systematic errors in the inferred cosmological parameters.

\item {\it Pre-survey critical decisions}: Spectroscopic surveys differ crucially from
imaging surveys, because decisions need to be made about the target sample before the
spectroscopic survey is started.  Target pre-selection influences the possible instrument
configurations and increases the importance of modeling the measurement process at early
stages before the data are collected.  Given limited time and resources to perform a survey,
it is likely to be difficult to drastically alter survey strategy at late stages to
alleviate systematic errors arising from target selection.

\item {\it Heritage:} Mapping large-scale structure with wide-field spectroscopic surveys is
a mature field. As a result, there exist numerous tools and methods that have been developed
for their exploitation. The incorporation of these resources is highly desirable, but also
challenging. Many tools originate in heterogeneous code bases, making it difficult to
integrate them into a coherent framework without significant modification .

\end{itemize}

\subsection{Simulation Requirements}\label{sec:key_ingredients}

\begin{figure*}[!ht] \hspace{-10mm} \centering \vspace{0mm}
\includegraphics[width=145mm]{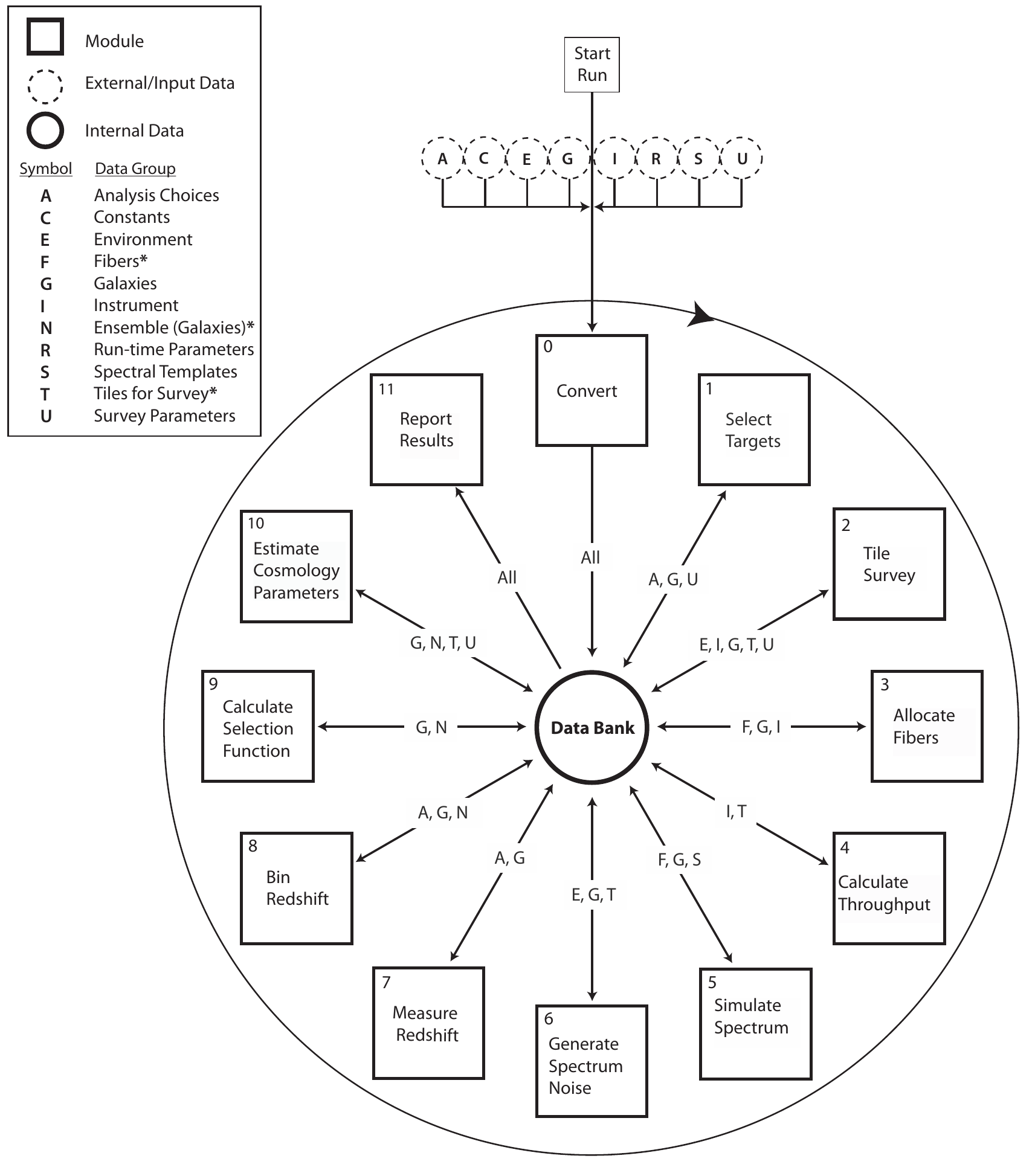} \vspace{-2mm} \caption{The
workflow of the \spokes\ pipeline depicts the data management, the modules and operations
needed to simulate a spectroscopic survey and the sequence of those operations.  Data are
represented by circles and discussed in detail in \S\ref{sec:data}: external input data
(dashed) are further delineated in Table~\ref{table:parameters}; internal data (solid) are
held in the central \databank\ in the native \spokes\ format.  The data are split into data
groups to simplify provenance and access. Each data group is represented by a symbol, as
shown in the legend, and the data groups marked with `{\bf *}' are created within the
pipeline and not ingested from an external source.  The modules (squares) access (arrows)
data from the \databank, but otherwise do not interact.  The data groups that are used or
created by a given module are listed on that module's access arrow. In addition, we describe
the specific data elements used and created by each module in \S\ref{sec:modules}. The
wheel-like format of the pipeline shows the data management scheme, the independent nature
of the modules, and the cyclic nature of the pipeline's execution. Once the results have
been analyzed, the user can update parameters or data sets and re-run the pipeline in an
effort to meet the science goal; this can be done by hand or in an automated fashion.
}
  \label{fig:flow}
\end{figure*}

Extensive simulations are a key element in meeting these challenges, because they can be
used to design and validate an experiment.  Furthermore, simulations will play an important
role in developing the data processing framework for the scientific exploitation of the
surveys.  Several ingredients are required of the pipeline for it to be used for these
purposes.

First, to fully predict the results of a survey and to contend with complex or non-linear
coupling of errors, the simulations should track all steps of the experiment. This points
toward the need for an {\it end-to-end} simulation framework. It would start with mock
galaxy catalogs specific to input cosmological parameters, perform all the elements of the
experiment and analysis, and then end with constraints on those cosmological
parameters---all performed in a single run of the simulation.

The architecture will need to be fully {\it integrated}, such that all the functions
communicate with the rest of the pipeline components through the same mechanism---ensuring
data and logic to be tracked precisely (i.e., provenance).  In particular, the different
simulation functions should pass parameters and data consistently and clearly from one
function to the next.

The simulation framework should allow for a high level of {\it reproducibility}: one should
be able to produce identical results with identical inputs. This requires careful and
comprehensive book-keeping of all relevant parameters.  For example, stochasticity needs to
be controlled by saving the value of random seeds for each calculation, such as for the
generation of noise in target spectra. This fine level of provenance is critical for
systematic optimization of experiment parameters.

The framework must nevertheless be sufficiently {\it flexible} to permit the ingestion of
new heterogenous functions and the modification of current functions. In addition, some
operations of the experiment or analysis of the data will be time-consuming or memory-heavy.
Therefore, the pipeline will need to accommodate a range of execution modes---on the one
hand accomodate high-speed, high-efficiency runs and on the other permit
computationally-intense (e.g., image-level) calculations, which may require parallelization.
The simulation pipeline should thus be sufficiently flexible to accomodate a {\it high
dynamic range} in detail (of what is simulated) and in parallelization.

The run time of the pipeline will depend on the run mode --- i.e., the level of detail
simulated. There should be a `minimal' mode that can run relatively precisely to recover the
correct output of the experiment, while running fast enough to allow for a large number of
iterations to explore a large space of input parameters.  Exploration of this parameter
space can give us a better understanding of the sensitivity of the experiment to systematic
effects and variation in experiment parameters \citep[e.g.,][]{Refregier:2013un}.

%-----------------------------------------------------------------------
\section{The \spokes\ Facility} \label{sec:solution}
%-----------------------------------------------------------------------

\begin{figure}[!htb] \hspace{-4.5mm}
\includegraphics[width=95mm]{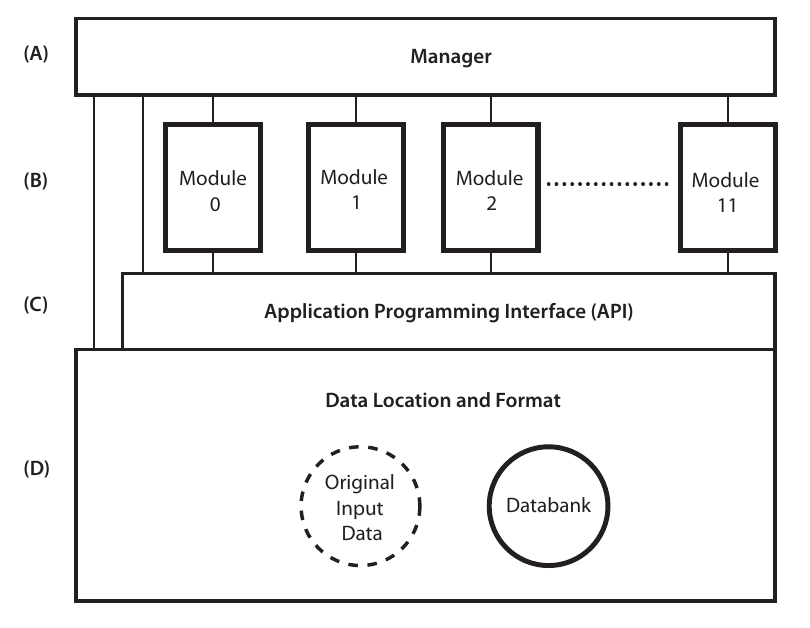}
\caption{Architecture of the \spokes~pipeline. The \gluevar\ (section A) connects the
multiple elements of the pipeline, uses the API to access data, and is the main interface
for the user: it is the point where the user selects the modules to use and sets the
simulation run parameters.  The modules (B) read and write data via the API, but they are
otherwise independent of one another.  The API (C) handles data access throughout the
pipeline, starting with conversion of the ``original input data'' (dashed circle)  to the
\spokes\  ``\databank'' format (solid circle) and storage in the central \databank\ (D). }
\label{fig:architecture} \end{figure}

The \spokes\ simulation facility is designed to meet the above requirements for simulation
pipelines in wide-field spectroscopic surveys.  There are two principal layers in the
\spokes\ pipeline. The algorithmic layer (shown in Fig.~\ref{fig:flow}) governs the aspects
related to  experiment planning and science analysis. The infrastructure layer (shown in
Fig.~\ref{fig:architecture}) contains the novel computing elements that allow the pipeline
to meet requirements of next-generation simulations.

Fig.~\ref{fig:flow} shows the pipeline in an algorithmic context: sequentially and in
clock-wise order, each function takes data and parameters from the \databank\ and creates
new data to be used later in the pipeline. The first task of the pipeline is performed by
Module 0, which imports multiple sets of heterogeneous data into a \databank\ (see
\S\ref{sec:data} and Table~\ref{table:parameters} for details). The pipeline then proceeds
to perform the functions of a spectroscopic survey. It selects targets from mock photometric
catalogs, then performs survey and fiber allocation operations, measures spectroscopic
redshifts, and derives cosmological constraints (Modules 1-10; see \S\ref{sec:modules}). At
the conclusion of computations, Module 11 produces a summary report that includes diagnostic
plots and statistics from a single run through the pipeline. The pipeline can also be
directed to run cyclically---either traversing a pre-determined space of input parameters or
searching through the parameter space until some metric is optimized.

 Each module accesses only the data it needs from the \databank, which simplifies the
 interfaces between modules and makes them highly independent of one another. Note that the
 only interaction between modules occurs via the exchange of data with the \databank.  The
 data groups that are read, used or written by each module are shown along the arrows
 between the module and the \databank. The legend in Fig.~\ref{fig:flow} provides a symbol
 for each data group, and the list in \S\ref{sec:data} describes the contents of each group.
 We also note specific data products that are used or created by each module in their
 respective descriptions.
%}{2M3}

The data management architecture of the pipeline is shown in Fig.~\ref{fig:architecture}.
The {\it Manager} is the top layer (A) of the infrastructure, and it organizes the interplay
among the elements of the pipeline and manages execution.  The second layer (B) shows the
{\it Modules}: they do not talk directly to one another, but access the data via the {\it
Application Programming Interface} (C). The {\it Data Location and Format} layer (D) contain
the \spokes\ data products, including the original input data, user-specified parameters and
all data created throughout the pipeline. Below, we discuss each layer in detail.

\subsection{\gluevar\ }\label{sec:glue} The topmost layer of the infrastructure, the
\gluevar\ (Fig.~\ref{fig:architecture}A),  combines and coordinates all components of the
pipeline. It is responsible for merging configuration files, placing all data and parameters
into the data bank, managing the modules, and executing the pipeline.

The user first sets up the simulated experiment: a master parameter file contains the
experiment parameters that will be used by the modules (e.g., Table~\ref{table:parameters}),
and a task scheduler contains the modules to be used, the sequence in which they are run,
and the compiler and executable that will run the modules. The \gluevar\ parses the
parameter and scheduler files and executes each module in order. To validate results
throughout the pipeline run, the \gluevar\ also performs module-specific quality assurance
tests (see \S\ref{sec:qualityassurance}).

% ===================
% Modules
% ===================
\subsection{Modules}
\label{sec:modules}

Pipeline operations are broken into discrete {\it modules} (Fig.~\ref{fig:architecture}B),
which are executed by the Manager in the order specified by the user. Each module reads the
particular data it requires from the \databank\ (Fig.~\ref{fig:architecture}D), performs a
specific task, and then writes new data to the \databank. The modules are independent of one
another: the only interaction between them occurs through the \databank. Any module can be
replaced without disrupting the rest of the pipeline, as long as the data needed by each
module is ingested at the beginning of the pipeline (Module 0, \S\ref{app:sec:convert}) or
provided by a preceding module in the pipeline. The following describes the purpose and
algorithm of each module (shown in Fig.~\ref{fig:flow}), as implemented to forecast DESpec.

% ------------------------
% \item[Module 0]
% ------------------------
\subsubsection{Module 0: Convert}
\label{app:sec:convert}

The starting point for the computation is catalog-level galaxy data (originating from a
precursor imaging survey) and user-defined parameters (describing the design of the
instrument and survey). This module imports the parameters and one or more catalogs into the
pipeline's central \databank, which is then used as the sole data repository by the rest of
the modules. The variables and the format of the \databank\ are described in detail in
\S\ref{sec:data}, and our specific choices for parameters are described in
Table~\ref{table:parameters}.

% ------------------------
% \item[Module 1]
% ------------------------
\subsubsection{Module 1: Select Targets}
\label{app:sec:target_selection}

This module selects targets for spectroscopic observation from the photometric catalog of
galaxies in the \databank\ using user-defined parameters for color and magnitude cuts.
Target selection is performed separately for the large red galaxies (LRGs) and for the
emission-line galaxies (ELGs). We assume that all DES filter bands ({\it grizY}) are
available in the catalog photometry, but we do not use the {\it Y} filter.

For the LRG cuts, the target selection criteria are $z<22$ and $(r-z)>1.5$. As shown in
\cite{Abdalla:2012tm}, these cuts are expected to provide a relatively flat redshift
distribution for LRGs over the redshift range $0.5<z<1$. For ELGs, the selection criteria
are $i<23.5$; $0.1< (r-i) <1.3$; and $-0.2<(g-r)<0.3$. This is similar to the ELG selection
cuts described in \cite{Schlegel:2011vk}.

For the implementation used in this work, the circular field of view has a radius of 1.1 deg
and an area of 3.8 deg$^{2}$. Four thousand fibers are distributed within a regular hexagon,
itself inscribed in the circular field of view. The hexagonal field of view then has an area
of 3.14 deg$^{2}$. If there are two passes on a given tile (patch of sky; see Module 2),
there is then an effective fiber density of $2*4000/3.14 \sim 2550/$deg$^{2}$ for each tile.
We seek to use the fibers efficiently both by placing them on target galaxies, and by
keeping some available for measurements of the sky background and for community projects. To
satisfy these constraints, and to dynamically choose the fraction of ELGs and LRGs, we apply
a random spatial sampling and elect to keep 100\% of ELGs and 18\% of LRGs. The output is a
set of all galaxies flagged for spectroscopic observation.

% ------------------------
% \item[Module 2]
% ------------------------
\subsubsection{Module 2: Tile Survey}
\label{app:sec:survey_strategy}

This module implements the survey strategy by tiling the instrument field of view across a
user-specified sky region, while optimizing observations for simulated environmental and sky
conditions. The module has two main functions---the Planner and the Scheduler. They take as
inputs the positions of targeted galaxies, as well as parameters of the fiber positioner
(e.g., fiber arrangement and tile shape) and of the survey strategy (e.g., exposure time,
area and number of observation passes per tile).

The Planner uses a survey mask and a hexagonal tiling pattern to geometrically
optimize the survey area coverage: a mask designates the region of sky to cover, and the
hexagonal tiles provide closely packed observed fields.

The result is a list of tiles and
their celestial coordinates. The Scheduler uses the list from the Planner to select
observation dates and times for each tile within the survey area, optimizing observing night
efficiency by accounting for sky brightness and airmass. The scheduling is constrained by a
set of user-defined parameters---observing dates, the exposure time, an airmass limit and a
sky brightness limit. At the beginning of a night, the scheduler first identifies the tiles
that are visible. The visible tile that is within the airmass limit and that has the
smallest sky brightness is selected for observation. If no tile meets these criteria, then
the Scheduler increments the time, and checks again. The scheduler ascends through the
visible tiles by their Right Ascension, repeating the tile-selection process.

We model the sky brightness by estimating the moon brightness and visibility, the zodiacal
light, and the airglow (and for each location of the moon in the sky during a night). The
model does not account for clouds. The seeing is modeled via a Gaussian distribution with a
mean of 1.0 and standard deviation of 0.25: this distribution is sampled to provide each
targeted observing tile with a seeing value. The seeing distribution does not affect how the
pipeline is operated, but it could affect how the pipeline performs: in this implementation of SPOKES,
there is an upper limit to the seeing for exposures that may be acquired. This
affects the number of galaxies (with their redshifts) that may be used for analysis, which
then affects the final cosmological constraints: e.g., if the seeing distribution is skewed high, then fewer
exposures may be acquired during the survey, potentially increasing the statistical
errors in the cosmological constraints (Module 11).

The tiles are not permitted to overlap in this implementation, and tiling does not
account for relative target densities. The latter issue is addressed grossly in the target
selection module (Module 1), where random spatial sampling of each target populaion --- ELG
and LRG --- selects for a total target density that correlates with the experiment's fiber
density; some fibers may be kept free in order to accomodate community studies and measuring
sky backgrounds. The choices made for the survey tiling are closely related to those of the
fiber assignment: augmentations of the tiling and fiber assignment algorithms can be
implemented, and one should take care to keep separate the two modules or create a new
module that combines the two.

This module outputs a list of scheduled tiles for each night during the observing run. The
list includes the time of observation, sky brightness, seeing, airmass, celestial
coordinates and unique identification number for each tile scheduled for observation.
Additionally, we flag galaxies targeted in Module 1 that fall within a tile pointing: this
increases computational efficiency for fiber allocation in Module 3, because it sets the
only galaxies that will need to be read in for that Module.

% ------------------------
% \item[Module 3]
% ------------------------
\subsubsection{Module 3: Allocate Fibers}
\label{app:sec:allocate_fibers}

This module matches fibers to positions in the focal plane of targeted galaxies (see Module
1) for each tile scheduled in the survey (Module 2). This allocation process is constrained
by instrument specifications (i.e., fiber patrol radius, fiber arrangement, field shape) of
an automated fiber-positioning system \citep[e.g., Mohawk, OzPoz, and
Hydra;][respectively]{2012SPIE.8446E..4WS, 2000SPIE.4008..914G,Barden:1995by}. DESpec is
designed with a Mohawk positioning system, which employs tilting spines to position the
fibers in the focal plane.

The fiber allocation module takes as inputs 1) the sky positions for each tile scheduled by
the Survey Strategy module; 2) the positions for the target galaxies produced by the Target
Selection module;  and 3) all the numbers describing the fibers---fiber diameter, the number
of fibers along the diameter of the hexagonal tile, the hexagon radius in degrees, and the
fiber patrol radius. The patrol radius describes the distance that a fiber can travel from
its rest position on the focal plane. These are the only relevant parameters for the current
implementation. DESpec's usage of the Echidna-based Mohawk fiber positioner is described in
more detail in Section 4B of \cite{Abdalla:2012tm}.

All the fibers begin in a fiducial spatial configuration---arranged in a hexagonal pattern,
all equidistant from one another and each with a unique identification number. In this
configuration, each target in the sky can be reached at least (most) by three (four) fibers.

The algorithm first chooses the galaxies that are going to be matched to a fiber by
prioritizing galaxies according to their local galaxy density: this scheme gives priority to
galaxies in crowded regions. We calculate the local galaxy density by estimating the number
of target galaxies within one patrol radius, $n_{p}$, of each galaxy. For each fiber, we
calculate a list of galaxies that can be reached. This list is then ranked in descending
order of $n_{p}$. For each fiber, the galaxy with the highest $n_{p}$ is allocated first.

When the algorithm attempts to match fibers to galaxies, the fiber movement paths can
intersect. In that scenario, it is not possible for all of the colliding fibers to reach
their matched galaxies on the focal plane. In the event of such a fiber collision, the two
or more colliding fibers are reset to their positions in the fiducial configuration. The
allocation process then begins again, choosing different galaxies to match to fibers, in
order to avoid the same collisions. The cycle iterates until the number of fiber collisions
cannot be decreased or the number of collisions is zero. When the fiber collisions cannot be
decreased, the fiber remains unmatched for this tile.

The output is a list of galaxies, each flagged with a unique identification number of the
fiber that will be used to observe the galaxy. The algorithm has been made
public\footnote{\url{https://github.com/forero/FiberAllocation/blob/master/text/note.pdf}}
and is described in more detail in \citep{2014SPIE.9150E..23S}.

% ------------------------
% \item[Module 4]
% ------------------------
\subsubsection{Module 4: Throughput}
\label{app:sec:throughput}

This module calculates the total optical transmission efficiency as a function of wavelength
for the principal elements in the light path of the instrument. Before running the pipeline,
we estimate separately the throughputs of the main contributors to light loss---optical
elements in the telescope barrel, fibers, fiber positioner, and spectrograph---and then
ingest them into the pipeline in Module 0.

To model the barrel optics, we employ ZEMAX\footnote{\url{https://www.zemax.com}}. For the
fiber positioner, the main contribution to throughput loss comes from the effect of the
fiber aperture (assumed to be circular). When a Mohawk spine undergoes significant tilt,
impending light from the barrel optics is apodized due to focal ratio degradation (FRD),
which causes light exiting the fiber to overfill the spectrograph collimator. We model the
input to the fiber for a variety of galaxy types as a convolution of four two-dimensional
galaxy radial luminosity profiles---Moffat (Beta), Gaussian, deVaucoleur and exponential.
The throughput of the fiber aperture is estimated as the fraction of the input beam that is
captured by the collimator, assuming the maximum FRD. This is discussed in more detail in
\cite{2012SPIE.8446E..4WS}.

For attenuation along the fiber, we use an estimate for a broadband optical fiber that is 50
meters in length and 100 microns in diameter \citep{Abdalla:2012tm,2012SPIE.8446E..56M}.
DESpec's lower wavelength limit is 480 nm. Experiments that seek to capture light below
$\sim450$ nm will contend with additional throughput losses from the fibers and atmosphere
for light near the short-wavelength end of the spectrum. First, within the fiber, Rayleigh
scattering caused by microscopic variations in the propagating medium's index of refraction
will increase attenuation. Second, wavelength-dependent refraction causes atmospheric
dispersion, which also depends on the pointing of the telescope; this could could be
mitigated with an atmospheric dispersion corrector. Finally, differential refraction can
cause the throughput and signal-to-noise to vary across the field
\citep{1989PASP..101.1046D}. If these issues are not addressed in the survey strategy and
in the instrument design, themselves, then the throughput modeling must account for the
effects incurred by these processes.

For the spectrograph, the most important components are the dispersive element and the CCD
detector. We model the dispersive element as a volume-phase Holographic (VPH) grating. The
spectral efficiency of the grating is estimated via VPH
GSolver\footnote{\url{http://www.gsolver.com/}}, which finds solutions to the general
diffraction grating problem for periodic grating structures.  In the spectrograph, detectors
are based on the Dark Energy Camera (DECam) CCDs, and we use measurements of the CCD
throughput as our model \citep{2010SPIE.7735E..5CK}.

The Throughput module combines the results from the individual elements into the complete
transmission efficiency (as a function of wavelength in Angstroms), except for the effect of
the atmosphere. While the atmospheric power affects this calculation, it is implemented in
Module 6 (\S\ref{app:sec:genspecnoise}) for computational efficiency.

% ------------------------
% \item[Module 5]
% ------------------------
\subsubsection{Module 5: Simulate Spectrum}
\label{app:sec:spectrum}

This module constructs models of the intrinsic rest-frame and of the observed-frame spectral
energy distributions for each galaxy that has been scheduled for targeting. Each rest-frame
spectrum is a linear combination of five {\it kcorrect} templates
\citep{2003ApJ...592..819B}, which are themselves derived via the non-negative matrix
factorization technique \citep{2007AJ....133..734B}. The templates are empirically-derived
from known galaxy types. A coefficient for each template is derived from the photometry of
the galaxy, and it determines the amount of that template's contribution to the total
spectrum for a galaxy. The choice of template spectra is described in
\cite{2012arXiv1207.3347C}. The outputs from this module are 1) a rest-frame spectrum and 2)
a wavelength-redshifted and flux-dimmed spectrum for each galaxy. The wavelengths are in
units of Angstroms, and the fluxes are in units of $ergs\, cm^{-2} s^{-1} \AA^{-1}$.

% ------------------------
% \item[Module 6]
% ------------------------
\subsubsection{Module 6: Generate Spectrum Noise}
\label{app:sec:genspecnoise}

In this module, the transmission throughput and simulated spectra---generated in Module 4
and Module 5, respectively---are used to produce a complete noise spectrum that also
includes photon shot noise, spectrograph CCD read noise and noise from the atmosphere
(extinction and sky background). Atmospheric absorption comes from the Palomar sky
extinction model (from B. Oke and J. Gunn), and the atmospheric emission from optical sky
background models from Gemini\footnote{Sky spectrum obtained from
\url{http://www.gemini.edu/sciops/ObsProcess/obsConstraints/atm-models/skybg\_50\_10.dat}}.
The atmospheric emission model is an estimate of the dark optical sky at airmass of 1.0 and
7th day illumination. The atmospheric spectra are adjusted for the airmass of the tile in
which the galaxy is observed (set by Module 2, \S\ref{app:sec:survey_strategy}), but not for
the illumination or position of the moon. More details of the reconstruction and noise
generation can be found in Appendix A2 of \cite{2012arXiv1207.3347C}. The output of this
module is a noise spectrum for each galaxy.

% ------------------------
% \item[Module 7]
% ------------------------
\subsubsection{Module 7: Measure Redshift}
\label{app:sec:measure_redshift}

This module measures the spectroscopic redshift, $z_{{\rm spec}}$, of the galaxies from
observed spectra. The observed spectra are the combination of the observed-frame spectra and
the noise --- generated in Modules 5 and 6, respectively.

To measure the redshift of a galaxy, we perform a chi-square minimization between the mock
observed galaxy spectrum and a set of model spectra. The chi-square minimization is
performed on a grid of redshift values: for each redshift, we linearly optimize the five
coefficients of the model spectrum (see \S \ref{app:sec:spectrum}), which are constrained to
be positive-definite. The best-fit redshift is taken from the redshifted combination of
linearly optimized models that best matches the observed spectrum. The output is a list of
best-fit redshifts, along with chi-square values (used to judge the quality of fit) for all
the observed galaxies.

% ------------------------
% \item[Module 8]
% ------------------------
\subsubsection{Module 8: Bin Redshift}
\label{app:sec:bin_redshift}

This module distributes the galaxies into bins of spectroscopic redshift (measured in Module
7), according to a user-defined parameter for the number of bins---chosen to be five for
comparison with the DESpec white paper. The output of this module is a one-dimensional
distribution to be used in Module 10 for a tomographic power spectrum analysis.

% ------------------------
% \item[Module 9]
% ------------------------
\subsubsection{Module 9: Calculate Selection Function}
\label{app:sec:selection}

This module calculates the selection function in space (Right Ascension and Declination) and
redshift of the observed spectroscopic galaxy catalog. We use the spectroscopic redshift
distribution from Module 8 and the true redshifts of galaxies from the input galaxy catalog.

In order to estimate cosmological parameters, we must use the true galaxy distribution, not
the spectroscopically observed one: the true redshifts determine the galaxy positions, and
therefore, the theoretically expected correlation functions. Through the relationship
between the spectroscopic redshift and the true redshift, the true galaxy distribution can
be computed for each spectroscopic redshift bin obtained in Module 8.

The outputs of this module are a redshift selection function---the distribution in true
redshift for each of the spectroscopic redshift bins---and the fraction of sky that has been
observed completely.

 ------------------------
% \item[Module 10]
% ------------------------
\subsubsection{Module 10: Estimation of Cosmological Parameters}
\label{app:sec:estimate_params}

The last computational step of the pipeline forecasts the cosmology-constraining power of a
given survey configuration by analyzing the catalog of galaxies observed in this pipeline.
We perform a power spectrum analysis for cosmological parameter estimation tomographically
in redshift.

This function uses a Fisher Matrix analysis of the number density distribution of the
galaxies from Module 8, in combination with the selection function from Module 9, to
estimate constraints on cosmological parameters. The Fisher analysis uses the
redshift-binned spherical harmonic power spectrum $C_l^{ij}$. We include cosmic variance and
galaxy shot noise, as well as redshift space distortions, but we neglect galaxy bias. We
vary a set of seven $\Lambda$CDM parameters: $\{ h = 0.7,~ \Omega_{m} =
0.3,~\Omega_{\Lambda} = 0.7, ~w_{0} = -0.95, ~w_{a} = 0.0,~ n_{s}= 1.0,~\delta_{H} =
1843785.96 \}$. Details of the calculations are described in \cite{2014PhRvD..90f3515N}.

% ------------------------
% \item[Module 11]
% ------------------------
\subsubsection{Module 11: Report Results}
\label{app:sec:report_results}

\spokes\ achieves provenance by saving all the science data created by the run, as well as a
suite of computational diagnostic information. Based on those data, this module generates a
report that summarizes the run with figures for assessing the computational and science
performance. The report includes basic information about the run (e.g., module versions, the
order in which modules are called and a full list of the parameters used); statistics on
computational efficiency (memory usage and run times for each module); and statistics on
science performance (e.g., the dark energy figure of merit, the redshift distribution, all
spectra). The report is a LaTex-typeset PDF document that contains figures, tables and text
that can be used for in-depth analysis of a run, and thus for improving subsequent runs of
the pipeline.
%}}{2M8}

% ------------------------
% Cyclicity
% ------------------------
\subsubsection{Cyclicity}
\label{app:cyclicity}

The goal of \spokes's wheel-like framework is to enable informed and swift optimization of
the experiment for the chosen metric --- e.g., the dark energy figure of merit (FoM) for
DESpec. Upon assessment, this information can be used to update the modules, parameters and
catalog data to improve the metric results by trial-and-error, or to explore systematically
the dependence of the metric on these choices. The user can implement automated running of
the pipeline: a wrapper around the Manager could run the pipeline cyclically to traverse a
multi-dimensional grid of input parameters to explore the effect of each parameter on the
final computational and science outputs.

The large number of parameters may present challenges for optimizing the experiment either
by hand or by automated algorithm. There may indeed be multiple configurations that achieve
the same success (according to a science metric), and the result may depend on the methods
used for optimization. Markov Chain Monte Carlo (MCMC) methods (e.g., Metropolis Hastings)
have proven to be useful as techniques for optimization in high-dimensional modeling
problems. MCMC's or genetic algorithms could be used to run the pipeline all the way through
to a final optimized experiment configuration and a science metric, or they may simply
provide information for a decision of how to run the next iteration by hand.

In addition, one of the features of the SPOKES facility is to act as a test-bed for
hypotheses of how to design an experiment. That is, even if it is not feasible to
completely optimize an experiment to a specified science metric, SPOKES is a facility in
which to test experiment configurations---e.g., the interplay between various instruments in
the experiment.

%=====================
%=====================

\subsection{Application Programming Interface}
\label{sec:api}

The \spokes\ facility includes an Application Programming Interface (API), which has been
specifically designed to provide a simple, efficient, robust, and user-friendly interface
between the modules and the \databank\ (see Fig.~\ref{fig:architecture}B). The API handles
the access to the \databank\ and hides the internal workings of the data format (HDF5, see
\S\ref{sec:data}) from the module developers.  We simplify and abstract the data handling to
reduce the amount of code used for data access and so reduce possible sources of bugs.

The data handling is performed analogously to a Python dictionary, which has unique `keys'
to access a specific data element: this key is the full path name to the data element or
variable in the HDF5 file. In a read operation, the data are read from the \databank\ and
returned to the program, and in a write operation, the data are written to the HDF5 path.
The creation of new groups, data type handling, overwriting existing paths is done by the
API and hidden from the user. Currently, most Python and
Numpy\footnote{\url{http://www.numpy.org/}} data types are supported, except for an actual
Python dictionary itself, which is not needed. An example of reading and writing a data
element are, respectively, ${ wavelength\_range} = {
bank[`/Instrument/Spectrograph/wavelength\_range']}$ and ${
bank[`/gal/target\_selection\_flag']} = { target\_selection\_flag}$, where ``bank'' is the
\databank\ variable.

The API also includes a feature for querying a registered function, rather than accessing
static data directly. This is a useful feature for steps that would otherwise generate large
amounts of data. For example, each galaxy has a flux spectrum with a flux value for each
wavelength at which it is measured: given the resolution and wavelength range of modern
spectrometers, this can amount to over ~20,000 floating point values per galaxy.  Tracking
each floating point value through multiple modules for millions of galaxies is untenable.
Therefore, we use the registered function feature to generate the galaxy spectra as they are
needed rather than storing spectra of all the galaxies. Thus, Modules 5 and 6 register
functions in the \databank\ (through the API), which are called later in Module 7  (see
Fig.~\ref{fig:flow}). At that stage, a galaxy's intrinsic spectrum is generated, the noise
is added and the redshift is measured all in a single step without the information of the
galaxy wavelength bins being saved to the \databank. This allows information to be passed
between modules, without generating large static data, while preserving a strict separation
between the modules: the modules only interact with the API and never with each other.

When a function is registered, a Python module is created, along with a path that will be
used to call the function: the path for the function is accessed like a path that contains
data. The conventions for input, output, and point of usage for the registered function are
the same as for all other functions. Both parameters and data can be passed to the
registered function, and there is no limit on its complexity.  Additionally, the user must
take care to ensure that the input is available for the registered function, that it outputs
the correct data for the remaining functions, and that the registered function is called at
the intended juncture in the pipeline.

\subsection{Data Management}\label{sec:data}

A common approach for pipeline development is to employ a linear work flow with data passed
directly from one module to the next. This approach has a number of drawbacks. For instance,
data generated by a module early in the pipeline needs to be carried through intermediate
modules (that possibly do not act on the data) until they reach the module in which the data
are used. Such an approach is not efficient and reduces the independence of the modules:
i.e. if a later module is modified to use extra data that are produced by an earlier module,
all intermediate modules need to be modified as well.

A linear workflow is common in the early development stages of a pipeline, where the
frameworks have a tendency to be ad hoc and data transfer between functions and programs
occurs by hand---i.e., one programmer gives a result to another, who programs the next
function in pipeline. This is not common for more mature studies. For example, the LSST data
management system \citep{2015arXiv151207914J} and pipeline have a large-scale framework with
a much more complex workflow: the prototype included a `clipboard', with which modules
interact for reading and writing data as necessary \citep{Axelrod2010}; this has since been
replaced by in-memory Python variables\footnote{\url{https://docushare.lsstcorp.org/docushare/dsweb/ServicesLib/LDM-152/History}}.
The \spokes\ \databank\ is similar to this `clipboard', keeping all data from a pipeline run
to enable maximal provenance.

For the \spokes\ facility, we have adopted a centralized system for managing all data, as
shown in Fig.~\ref{fig:flow} and Fig.~\ref{fig:architecture}. This scheme separates the data
from the modules, so that a given module accesses only the data it needs and creates. As
well as maintaining modularity, such a centralized scheme more easily allows for provenance
and reproducibility.

For this central \databank\ model to work, the data formatting must be able to handle many
data types, scale efficiently to handle large amounts of data and be flexible enough to
store all data for a rapidly developed pipeline. After evaluating a number of possible
standards --- including Flexible Image Transport System (FITS)\footnote{
http://fits.gsfc.nasa.gov/} and relational databases - we adopted a solution based on the
Hierarchical Data Format (HDF5)\footnote{http://www.hdfgroup.org/HDF5}. The detailed
justification of this choice is given in \S\ref{sec:fits_hdf5}.

In an HDF5 file, the data are organized in unique paths, like a hard disk
filesystems---e.g., {\it /group/subgroup/dataset}: each data set resides in a `group' and
its `subgroup', which are named descriptively in \spokes\ to associate related data and
improve code readability. The data sets can be of a variety of data types, including arrays.

To take advantage of this organizational paradigm, our API (see \S\ref{sec:api}) provides
trivial access to any field or data group via this path and this path alone, regardless of
data type.  It provides modularity of data access: a module may use individual aspects of a
data group without having to read in all data. For example, a module can access galaxy
identification numbers and positions without reading all the other data that another module
might need.  We describe the data groups later in this section.

The \databank\ can be stored in any number of HDF5 files, as suitable to the application,
and there is no intrinsic limit to the file size.  The maximum data file size is dictated
merely by the available working memory and machine hard disk storage.   When HDF5 files are
sufficiently large, the file is split on the disk into multiple files that are logically
merged transparently.  \spokes\ supports parallel reading of HDF5 files (e.g., multiple
processes in a batch job can access simultaneously), but not parallel writing.

To estimate the data storage requirement we find that about 1 GB is needed to store a
complete \databank\ with $\sim$ 10M input galaxies. However, after photometric selection and
survey and fiber allocation operations, only about 10\% of these will be targets (see
Table~\ref{table:galaxies}), for which redshifts will be measured: this gives 1GB for 1M
redshifts.  If a Stage IV Dark Energy experiment aims to measure redshifts for up to 20M
targets, it will require 20GB for the total catalog, or 2GB for the targets alone.

The data groups within the \databank\ are partitioned according to module usage and related
information; these data groups are shown in the legend in Fig.~\ref{fig:flow}. All original
data from input, including parameters (e.g., telescope optics choices) and data (e.g.,
galaxies), are converted to the native data format and separated into $M$ data sets within
the \databank\ upon initialization of the pipeline, as shown in the base data layer of the
architecture in Fig.~\ref{fig:architecture}D. The data groups (in alphabetical order) are
described below

\begin{description}
\item[ Analysis Choices (\dg{A}) ] contains the information with which to
  specify the analysis methods---e.g, magnitude or color cuts for
  Target Selection (Module 1) and bin size for redshift binning
  (Module 8).
\item[ Constants (\dg{C}) ] holds physical constants and random seeds.
\item[ Environment (\dg{E})] contains the information regarding the
  atmosphere (absorption and emission spectra) and location (e.g.,
  elevation) at which the observations are taking place.
\item[ Fibers (\dg{F})] contains information  (e.g., location in
  focal plane) about the fibers that are assigned to galaxies.
\item[ Galaxies (\dg{G})] contains all galaxy data.
\item[ Instrument (\dg{I}) ]contains several subgroups representing the
  subsystems of the instrument---optics, fibers and
  spectrograph---each of which has several parameters.
\item[ Ensemble (\dg{N}) ]contains data on the galaxies as a
  collection---the redshift histogram, related cosmological
  constraints, etc.
\item[ Run-time Parameters (\dg{R})] are those which determine how the
  simulation will be run --- e.g., with or without parallel
  processing, or which simulation files are to be used.
\item[ Spectral Templates (\dg{S})] contain the eigentemplates used to
  reconstruct galaxy spectra.
\item[ Survey Tiles (\dg{T})] contains a set of tile information (sky
  position, airmass, time of observation, etc) and is used to link
  galaxies with the time and observation environment in which they
  were observed.
\item[ Survey Parameters (\dg{U})] holds the data necessary to run the
  survey, for example, exposure time per tile and region of the sky to be observed.
\end{description}

\begin{table*}[!htb]
  \caption{Selected baseline \spokes\ input parameters. These parameters have been chosen to match those proposed for DESpec.}
  \centering
\footnotesize
  \begin{tabular}{l c c c}
\hline\hline
\centering{Data Group} & Used by Module(s) &  Parameter &  Value \\ [0.5ex]
\hline  \hline
{\bf Analysis ({\it A}):  Target Selection} & & & \\
\hline
\multicolumn{1}{c}{LRG cuts} & 1 & $z$    & $<$ 22    \\
&1 & $(r-z)$  &  $> 1.5$  \\
\hline
\multicolumn{1}{c}{ELG cuts} &1 &$i$ &  $<$ 23.5  \\
&1 & $(r-i)$ & $> 0.1$ and $<1.3$  \\
&1 &  $(g-r)$  & $>-0.2$ and $<0.3$ \\
\hline
{\bf Environment ({\it E}): Atmosphere\textsuperscript{**}}  & & &  \\
\hline
&6 & Sky Background & Gemini Sky Models \\
&6 & Atmospheric Extinction & Palomar Extinction Curves \\
\hline
{\bf Instrument ({\it I}): Fibers} & & & \\
\hline
& 3& Number of Fibers & 4000 \\
& 2,3 & Fiber Arrangement & Hexagon \\
& 2,3 & FOV: Hexagon radius & 1.1 deg\\
& 2,3 & FOV: Hexagon area & 3.14 deg$^2$\\
\hline
{\bf Instrument ({\it I}): Telescope} & &  \\
\hline
&4 & Diameter & 4m \\
&4 & Optical Efficiency\textsuperscript{*} & $\sim 0.25$ \\
\hline
{\bf Instrument ({\it I}): Spectrograph} & & &\\
\hline
&6 & Read Noise & 5 e- \\
&6,7 & Wavelengths& [480, 1050] nm \\

\hline
{\bf Survey Plan ({\it U}) } & & & \\
\hline
&2& Exp. time & 1200 s  \\
&2,9,10 & Area & 5000 and 15000 deg$^2$  \\
& 2,3 & Passes per Tile & 2 \\
[1ex]
\hline
\hline
\end{tabular} \label{table:parameters}\\
The input variables to the pipeline and the values used in the
demonstration run of the \spokes\ pipeline.  Data groups and module
numbers coincide with those of Fig.~\ref{fig:flow}.  The 28 parametric
specifications shown here are necessary for running a spectroscopic
experiment. The table shows two types of target selection, ``LRG
cuts'' and ``ELG cuts,''
designed to preferentially target Luminous Red Galaxies and Emission-Line Galaxies, respectively.
\begin{flushleft}
\textsuperscript{*} \footnotesize{This value is the mean of the
  throughput spectrum.  See \ref{app:sec:throughput} for
  details of the throughput calculation. }\\
\textsuperscript{**} \footnotesize{see \ref{app:sec:spectrum}  for
  details}\\
\textsuperscript{N.B:} \footnotesize{The `Convert' module is not
  listed here, because all of the input parameters are ingested and
  placed into the \databank.}\\
\end{flushleft}
\end{table*}

\normalsize

\subsection{Quality Assurance}\label{sec:qualityassurance} An important aspect of building a
modular end-to-end simulation facility is to develop an integrated quality assurance (QA)
framework that will automatically detect possible problems with a given simulation run or
with an update to one of the modules. In \spokes, we are working towards building a
continuous integration process for performing tests at the unit and facility level. At
present, during pipeline runs, we have three discrete levels of tests and cross-checks for
QA.

QA first performs a series of basic logical cross-checks at the onset of each module. For
example, after fibers have been allocated to galaxies, the same function checks that each
galaxy has received some value flagging it as observed or not. It also checks that this
value is within the range of available fiber indices.

At the intermediate stage, there are a set of user-defined limits for specific properties of
the analysis, modules and galaxy catalog. When these limits are approached or exceeded, the
pipeline reports a warning.

\begin{figure*}[!htb]
\centering
%    \hspace{-0.5mm}
\includegraphics[width=150mm]{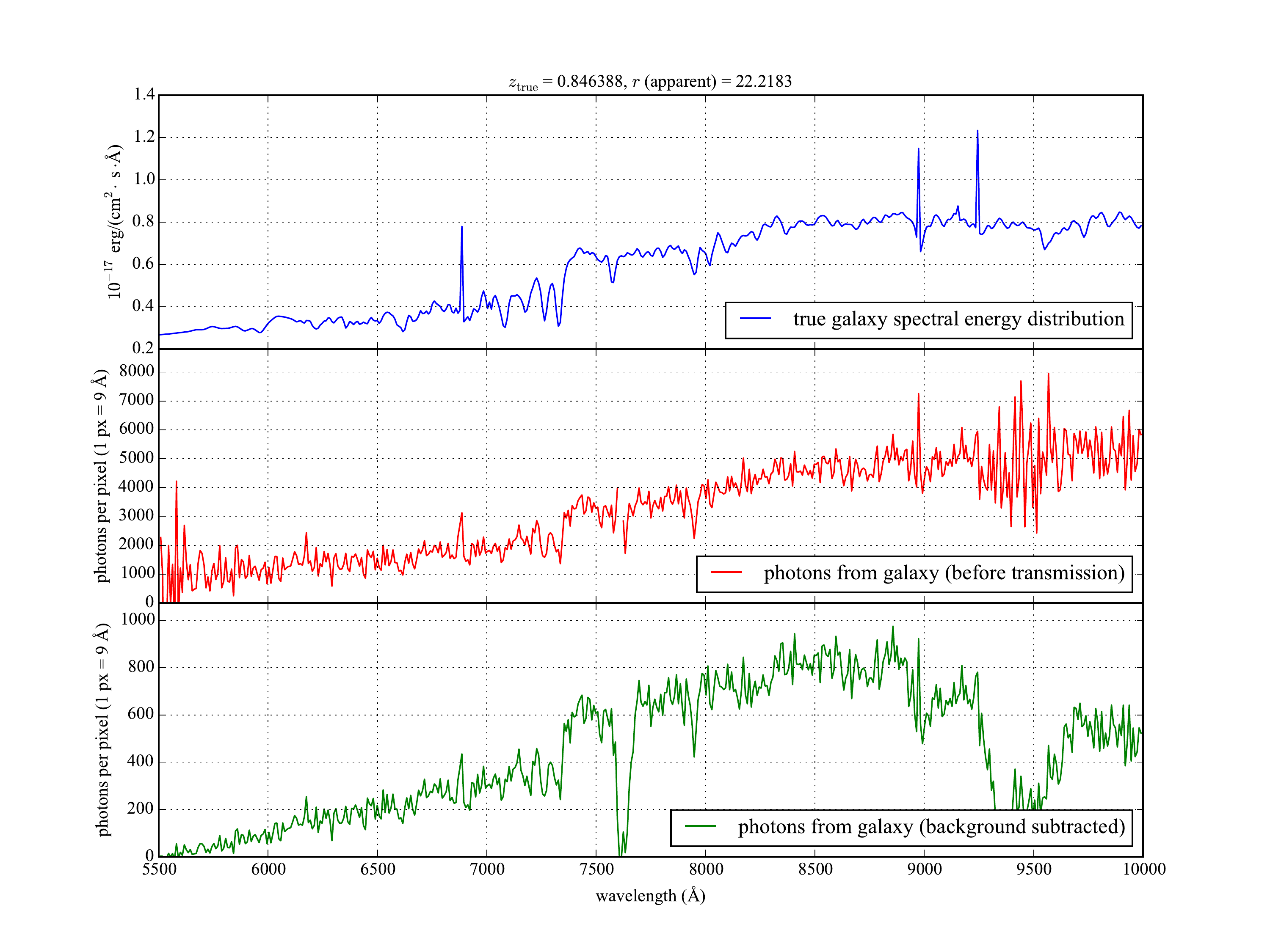}
\caption{Example of galaxy spectrum simulated by \spokes. Top panel: a noise-free galaxy
spectrum reconstructed from a set of templates for a galaxy at a redshift of $z = 0.93$.
Middle panel:  galaxy spectrum with noise computed from multiple sources---including Poisson
noise from photon counts, CCD readout noise and atmospheric sources---for an exposure time
of 1200 seconds. Bottom panel: galaxy spectrum with noise, but before being transmitted
through the atmosphere and telescope. More details of the spectrum and noise generation
process can be found in \S\ref{app:sec:spectrum}    and \S\ref{app:sec:genspecnoise}}
\label{fig:galaxy_spectrum_examples} \end{figure*}

Finally, the pipeline  provides scientific diagnostic figures and plots to check fidelity at
each step: e.g., Tile Survey (Module 2) produces a map of the fields observed.
Fig.~\ref{fig:galaxy_spectrum_examples} shows the spectral flux as a function of wavelength
for a single galaxy, with and without noise, respectively.  This pair of figures is produced
automatically for a randomly selected set of galaxies (the number of which is set by the
user) during each pipeline run. The user can then review the spectra to verify pipeline
accuracy at this juncture. This information is wrapped into an automatically generated final
report (by Module 11), with which the user can make assessments of the runs and input.
Another important aspect of QA is that we store sufficient information to ensure provenance,
so that each run can be reproduced at a later time.

%-----------------------------------------------------------------------
\section{Results}\label{sec:results}
%-----------------------------------------------------------------------

\subsection{Survey Baseline}

A large number of spectroscopic experiments are being planned or are underway.  To
illustrate the \spokes\ facility, we have decided to focus on DESpec \citep{Abdalla:2012tm},
an early concept that predates the upcoming DESI experiment and is representative of
next-generation spectroscopic surveys. DESpec was a  survey instrument designed for the
Blanco 4-meter telescope at the Cerro Tololo Inter-American Observatory (CTIO) in Chile.
Over the course of 1,000 nights, the planned survey would observe more than 20 million
galaxies and QSO's over 15,000 deg$^2$ of sky, with two passes for each patch of sky. The
data set would provide a three-dimensional map of the universe out to redshift $\sim2$. The
DESpec instrument design is primarily comprised of 10 spectrographs and an automated fiber
positioning system, Mohawk \citep{2012SPIE.8446E..4WS}, that can target the galaxies from a
precursor imaging survey, such as DES or LSST. Table~\ref{table:parameters} contains a
summary of the DESpec instrument and survey inputs that we consider as the baseline for our
calculations.

The DESpec design has a hexagon-shaped fiber plane with 4000 fibers of diameter 1.75 arcsec
(or 100 $\mu {\rm m}$) and pitch and patrol radius of $\sim 6 {\rm mm}$.  The spectrograph
has a read noise of 1-3 e- (assuming a gain of unity) and a wavelength range of 350 to 1050
nm. Further details of the spectrograph design can be found in the DESpec white paper
\citep{Abdalla:2012tm}.  We model the telescope optical efficiency based on ZEMAX modeling
of the Blanco optics: the average throughput across the wavelength range is $\sim 0.25$ (see
\S\ref{app:sec:throughput} for details of the throughput model).

\subsection{Pipeline Inputs}\label{sec:inputs}

In addition to the DESpec parameters summarized in Table~\ref{table:parameters}, one of the
key \spokes\ inputs is a galaxy catalog. For the results presented here we use a mock galaxy
catalog (described in detail in \S\ref{app:sec:simulation}), and we select experiment
parameters that correspond with the DESpec design concept described above. For this study,
we have used the mock galaxy catalogs based on the algorithm Adding Density Determined
GAlaxies to Lightcone Simulations (ADDGALS: Wechsler et al 2014 -- in prep., Busha et al
2014 -- in prep.). This algorithm attaches synthetic galaxies, including multiband
photometry, to dark matter particles in a lightcone output from a dark matter N-body
simulation and is designed to match the luminosities, colors, and clustering properties of
galaxies. The catalog used here was based on a single `Carmen' simulation run as part of the
LasDamas simulations \citep{2009AAS...21342506M}\footnote{Further details regarding the
simulations can be found at { http://lss.phy.vanderbilt.edu/lasdamas/simulations.html}}.
This simulation modeled a flat $\Lambda CDM$ universe with $\Omega_m = 0.25$ and $\sigma_8 =
0.8$ in a 1 Gpc/$h$ box with $1120^3$ particles. A 220-sq.-deg. light cone extending out to z =
1.33 was created by pasting together 40 snapshot outputs. For more details on the catalog
see \S\ref{app:sec:simulation}.

These simulations do not contain some classes of objects that would contaminate the target
selection---e.g., QSOs, low-mass stars, high-redshift galaxies.  Surveys, like the Galaxy
and Mass Assembly (GAMA)\footnote{\url{http://www.gama-survey.org/}} spectroscopic survey,
use a combination of optical and infra-red photometry to create cleaner samples
\citep[][Fig. 6]{2010MNRAS.404...86B}. The lack of contaminants will make the final results
for this implementation of \spokes\ better than a real survey. This work uses the same
simulations as the DESpec white paper, which allows for direct comparison with that work. In
this paper, we seek to compare to the DESpec white paper to show that such an analysis can
be performed more efficiently with the \spokes\ framework. Nevertheless, the purpose of the
\spokes\ framework is to be general enough to allow for the import of catalog data that has
these contaminants: to take advantage of an improved set of simulation data, the user would
be required to update the Target Selection module to work with this data.

The other key inputs to \spokes, such as atmospheric data on the sky background and
extinction, are taken from Gemini optical sky background models\footnote{derived from
\url{http://www.gemini.edu/sciops/telescopes-and-sites/observing-condition-constraints/optical-sky-background}}
and Palomar extinction
curves\footnote{\url{http://www.ing.iac.es/Astronomy/observing/manuals/html_manuals/tech_notes/tn065-100/small/palomar.tab}};
these are critical for constructing the noise model for each galaxy spectrum \citep[for
details see \S\ref{app:sec:spectrum} and ][]{2012arXiv1207.3347C}

\subsection{Science Performance}

To assess the science performance of the pipeline, we consider a typical \spokes\ simulation
run. Table~\ref{tab:gal_stat} traces the progress of the galaxy sample in one of the mock
catalog tiles through the pipeline steps.

The mock catalogs are stored in HEALPix \citep{2002ASPC..281..107G} cells, with $N_{side} =
8$, which corresponds to 53.7 deg$^2$ per cell. The initial mock catalog (ingested by Module
0) contains about $1.3 \times 10^5$ galaxies per square degree over about eight Healpix
cells. After magnitude and color cuts are applied, as part of  the target select/ion step
(Module 1), $4.8\times10^3$ galaxies per square degree remain. The vast majority of galaxies
are removed by the magnitude cut.

Once the tiling strategy has been defined (Module 2), the next key step for selecting
galaxies is the assignment of fibers in each tile (Module 3). Using the fiber allocation
scheme described in \S\ref{app:sec:allocate_fibers}, we find that the density of galaxies
assigned a fiber is 1750/deg$^2$. This can be compared with the maximal possible fiber
allocation density, 2550/deg$^2$ calculated in \S\ref{app:sec:target_selection}. This makes available
 $\sim800$ fiber per square degree (or $\sim 1250$ fibers per tile pass) for measurements of the sky background or for
community projects.

For each galaxy assigned to a fiber,  we simulate an intrinsic noise-free rest-frame
spectrum (Module 5), and a spectrum with noise from atmosphere and electronics (Module 6)
and with the signal reduced by the telescope throughput model (Module 4).  A typical
spectrum (with noise, without noise and before transmission through the atmosphere and
optics) is shown in Fig.~\ref{fig:galaxy_spectrum_examples}. The true or intrinsic galaxy
spectrum is shown in the top panel: note the prominent emission lines at $3727 \AA$ (OII),
$4861\AA$ (H$\beta$) and $5007\AA$ (OIII). In the noisy spectrum (bottom panel), the
prominence of these lines is degraded significantly---primarily by photon noise, the
dominant component of the overall noise. The effect of the statistical noise is seen clearly
at the red end of the spectrum, where larger errors are incurred due to the wavelength
dependence of the error. Finally, after the sky background has been subtracted and the
spectrum has been transmitted through the atmosphere and telescope (middle panel), telluric
features at $\sim 6900 \AA$ (${\rm O_2}$), $\sim 7600 \AA$ (${\rm O_2}$) and $\sim 9350 \AA$
(${\rm H_2 O}$) further degrade galaxy spectral features. The telluric features
could be removed with flux-calibrated standard stars. Telluric absorption features undergo
variation on time scales on the order of minutes, and they vary with the airmass of the
observation. This would require reserving a fiber for allocation of a standard star close in
time and airmass of an observed field. Alternatively, methods have recently been developed
to model telluric line spectra \citep[e.g.,][]{2010A&A...524A..11S, 2015arXiv151104641R}. In
either case, the process of removing features from each galaxy spectrum must be automated so
as not to require any human oversight.

Spectroscopic redshifts are then measured in Module 7. At this stage, we judge the quality
of the each of the fits using $\chi^2$ tests and reject galaxies with poor fits. From
Table~\ref{table:galaxies}, we see that this step removes about $\sim 7\%$ of the galaxies.
The only source of error in the spectra are related to photon counts (i.e., the error is
statistical, aside from systematic background flux sources and signal-to-noise reduction due
to transmission): there are no systematic errors in wavelength, so no emission or absorption
lines are systematically incorrect.

\begin{figure}[!ht]
%    \hspace{-10.5mm}
\centering
\includegraphics[width=95mm]{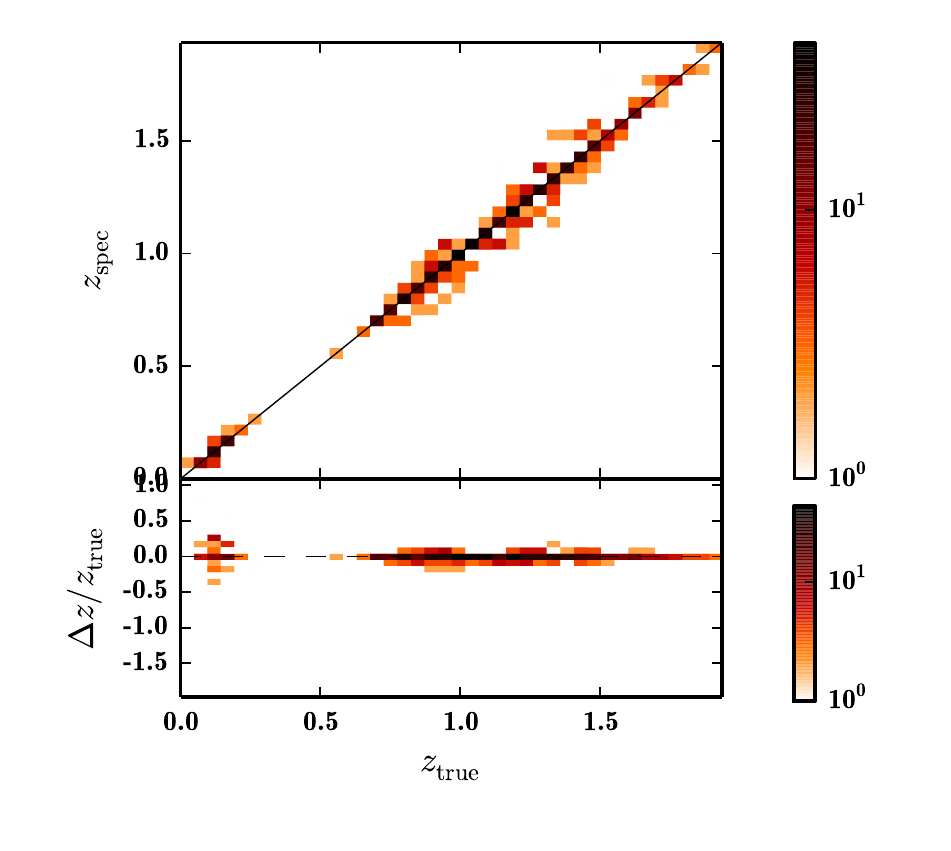}
\vspace{-12.5mm}
\caption{Comparison of true redshifts and the measured
    spectroscopic redshifts.  The upper panel shows a two-dimensional histogram of
  $z_{\rm spec}$ vs. $z_{\rm true}$, while the lower panel shows the fractional
  difference between redshift measures as a function of $z_{\rm
   true}$. In both cases, the solid black line corresponds to perfect recovery.
  The method used in \spokes\ to measure spectroscopic redshifts is described in
   \S\ref{app:sec:measure_redshift}. Note that only galaxies
     that meet the $\chi^2$ goodness-of-fit cut are included here. }
\label{fig:measure_redshift}
\end{figure}

\begin{figure}[!ht]
\includegraphics[width=90mm]{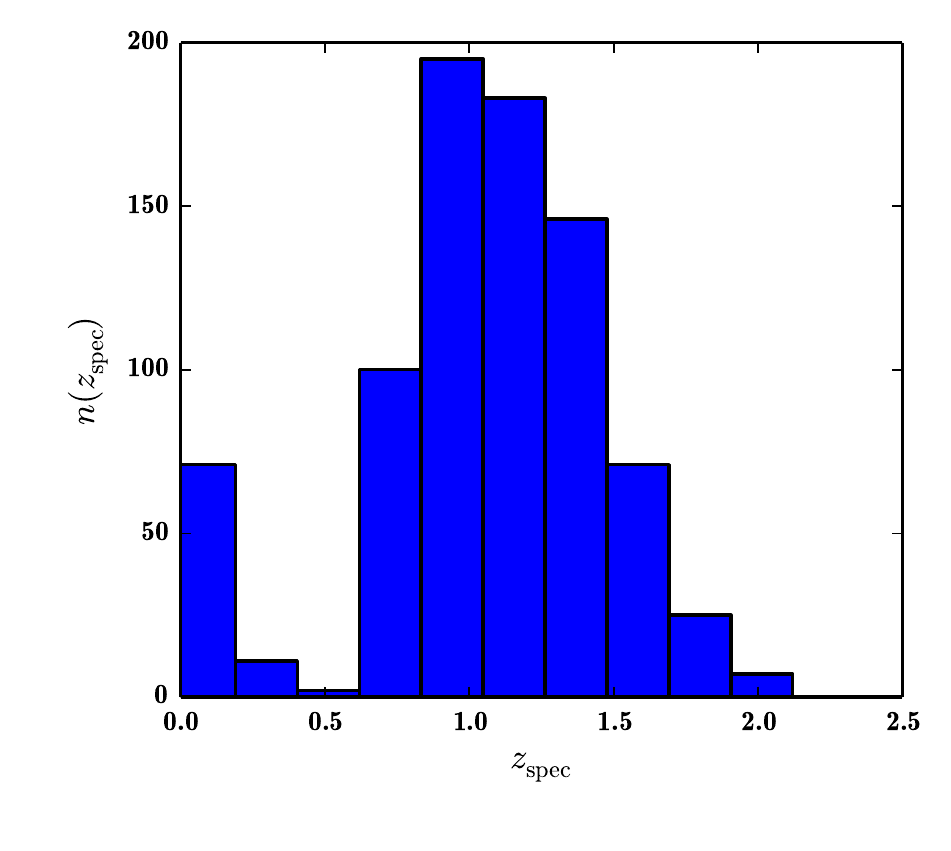}
\vspace{-7.5mm}
\caption{Redshift distribution ${\rm d} n/ {\rm d}z$ for the recovered
  spectroscopic sample. This is generated in Module 8 (see \S\ref{app:sec:bin_redshift}).}
\label{fig:bin_redshift}
\end{figure}

The resulting observed redshifts can be compared directly to the true redshifts coming from
the mock catalog, as shown in Fig.~\ref{fig:measure_redshift}. The figure shows that we are
able to recover the true input redshifts with small scatter. Note that the galaxies rejected
by the $\chi^2$ tests do not appear in the comparison between spectroscopic and true
redshifts. Different redshift measurement algorithms may obtain different success rates. We
also see that the underlying galaxies are not distributed evenly in redshift. For instance,
a gap around redshift of z=0.5 is prominent. The redshift distribution of galaxies in our
sample is driven by the cuts that we have made. In particular, we find that the $(r-z)$ cut
tends to select for galaxies at higher redshifts -- as was intended -- resulting in the
distribution shown in Fig.~\ref{fig:bin_redshift}. This figure shows the distribution of
spectroscopic redshifts, $n(z_{\rm spec})$, across bins measured in Module 8, and that our
galaxy sample peaks at a redshift $z\sim 1$ and extends out to redshifts of $z\sim2$.

\begin{table}
\caption{Galaxy statistics in a typical \spokes\ run\label{tab:gal_stat} }
\label{table:galaxies}
\centering
\footnotesize
\begin{tabular}{lll}
\hline\hline
module &  type & gals/deg$^{2}$  \\
\hline
0: convert &  all  & $1.3\times 10^5$   \\
1: select target  & LRG & $3.7\times 10^3$   \\
			&  ELG & $1.1\times 10^3$     \\
			& all 	 & $4.8\times 10^3$    \\
3: allocate fiber & LRG   & $6.5 \times 10^2$  \\
			& ELG   & $1.1 \times 10^3$  \\
			& all  & $1.7 \times 10^3$  \\
7: measure $z$ & LRG &  $5.9\times 10^2$  \\
			& ELG   & $1.0 \times 10^3$  \\
			& all  & $1.59 \times 10^3$  \\

\hline
\end{tabular}
\begin{flushleft}
\end{flushleft}
\end{table}

\begin{figure}[!ht]
\centering
\includegraphics[width=85mm]{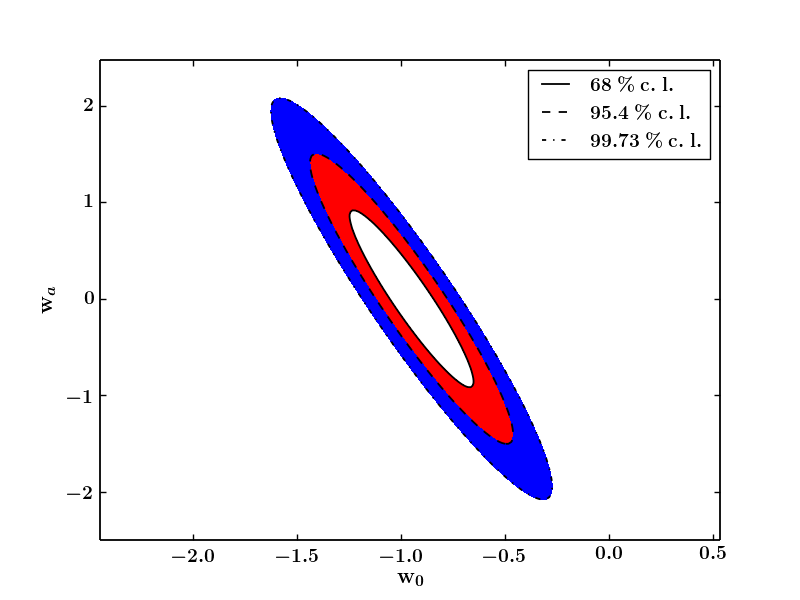}
%\vspace{10.0mm}
\caption{Confidence contours ($1, 2\ {\rm and}\ 3\sigma$ for white red
  and blue, respectively) for a joint estimate of dark energy
  parameters, $w_0$ and $w_a$.  See \S\ref{app:sec:estimate_params} for
  details on the calculation method. The results shown here include a WMAP9 prior that we construct from the publicly available MCMC chains.}
  \label{fig:estimate_cosmo_params}
\end{figure}

From the redshift distribution, we then measure a selection function, $W(z_{{\rm
spec}}|z_{{\rm true}})$ to obtain a distribution of true redshifts for each spectroscopic
redshift bin (Module 9).  This is then used with $n(z_{{\rm spec}})$ in the Fisher matrix
forecaster to compute confidence contours in the $w_0-w_a$ plane. Constraints on dark energy
equation of state parameters $w_0$ and $w_a$ are calculated in Module 10 (see
\S\ref{app:sec:estimate_params}) and the results, which include WMAP9 priors, are shown in
Fig.~\ref{fig:estimate_cosmo_params}.  The derived precision on these parameters is
consistent with that forecasted in the DESpec white paper \citep{Abdalla:2012tm}.

In forecasting DESpec, both the parameters of the experiment and galaxy catalogs remained
fixed. Through multiple runs of the pipeline, on a trial-and-error basis, we continuously
improved the modules. In particular, the computational efficiency of Modules 2, 3 and 7 were
increased dramatically. In future usage of the pipeline, we will seek to improve the
cosmological constraints through modification of the experiment parameters.

In addition to dark energy, one could study properties of (groups of) galaxies, such as the
luminosity function. This could replace the dark energy FoM as the metric for experiment
optimization, or it could be added to the current implementation and used in tandem. In
either case, one would write a new module to measure the galaxy properties of interest, and
instruct the Manager to call the module at the appropriate point in the pipeline. To develop
this enhancement, the user is required to develop new modules that incorporate galaxy
properties. This paper focuses on an implementation to forecast the FoM, where the
luminosity function is outside the scope. However, the \spokes\ framework is agnostic to the
choice of metric for the experiment, and the choice does not limit the framework at all.

\subsection{Computational Performance}

To evaluate the computational performance of the pipeline, we ran several benchmarks on a
laptop with an Intel i7 mobile processor with a clock rate of 2.2 Ghz and 8 GB memory. Some
of the modules, such as Module 2 (Tile Survey) and Module 7 (Measure Redshift), were
pararellised using eight threads on four cores.

\begin{figure}[!ht]
\centering
\includegraphics[width=90mm]{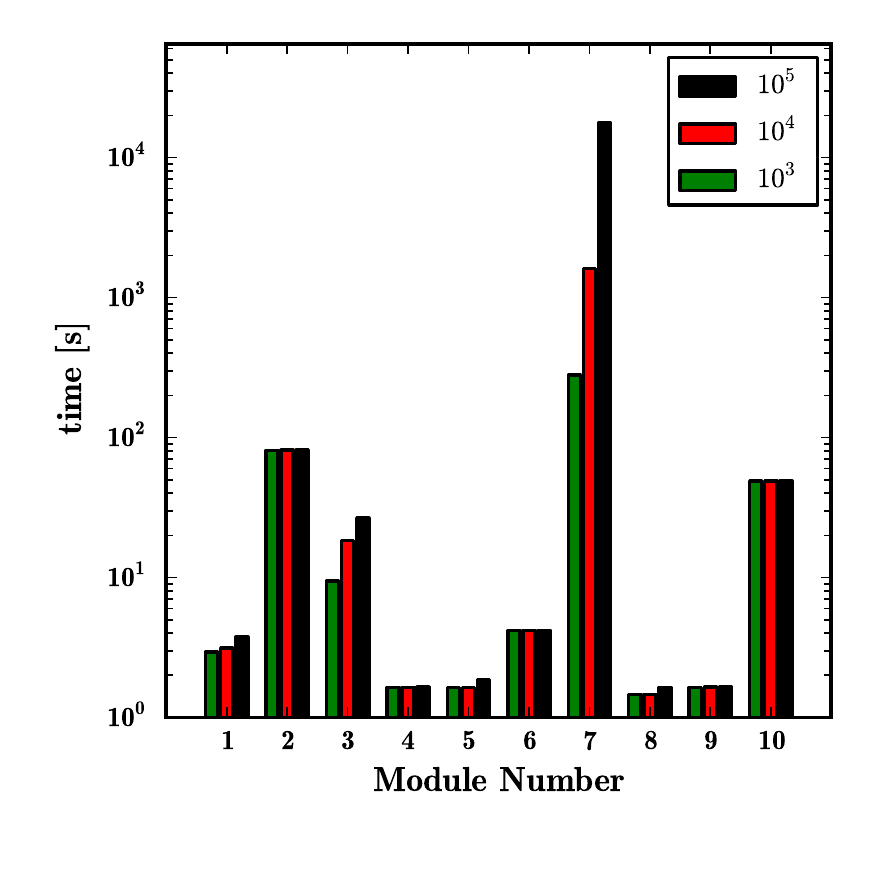}
\vspace{-7.5mm}
\caption{The time used by each module for 1k, 10k and 100k input
  galaxies.  }
\label{fig:module_timing}
\end{figure}

Fig.~\ref{fig:module_timing} shows the CPU time used by each module for different numbers of
input galaxies. As shown on the figure, a large fraction of the time is spent on the survey
tiling, fiber allocation, and redshift measurement---Modules 2, 3 and 7, respectively. The
latter is explained by the higher complexity of the redshift measurement process. The former
two are more surprising, and arises from the more detailed physics included in these modules
relative to others (e.g. weather modeling). These modules also have functions that are not
yet fully optimized (e.g. tiling, calculating fiber collisions, matching fibers to
galaxies). For example, the fiber allocation in Module 3 measures the distance between each
fiber and all galaxies multiple times: this is the principal time-consuming process. When
the count of galaxies per tile increases in this time-scaling test, the fiber allocation
module scales sub-linearly. If there are more tiles, but the same number of galaxies per
tile, then it will scale linearly.

Further optimisation reduces the CPU times of several of the modules. The parallelisation
scaling is seen in the dependence of the CPU times on the number of galaxies. It is close to
linear for Module 3 (Allocate Fibers) and Module 7 (Measure Redshift). Note also that for
all modules, execution time is dominated by CPU time, while I/O time is negligible with our
implementation.

\begin{figure}[!ht]
\centering
\includegraphics[width=90mm]{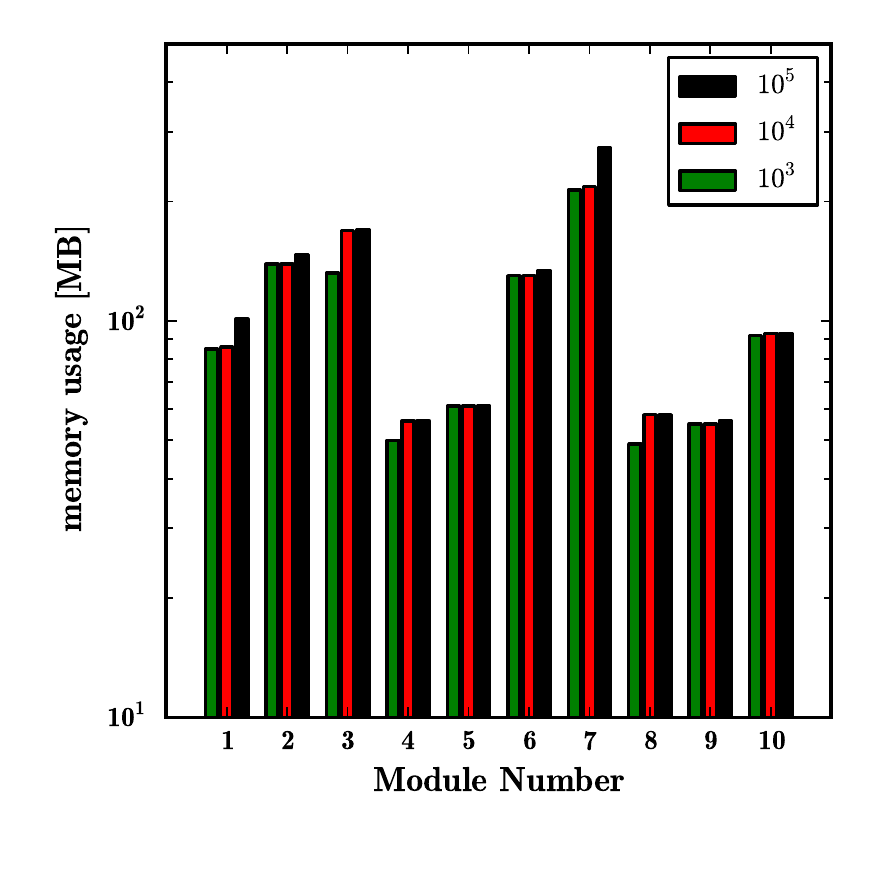}\
\vspace{-7.5mm}
\caption{The memory used by each module for 1k, 10k and 100k input
  galaxies.  }
\label{fig:module_memory}
\end{figure}

Fig.~\ref{fig:module_memory} shows the memory usage of each module. The highest memory
consumption takes place in Module 7 (Measure Redshift). The memory scales as a function of
number of parallel processes and galaxies since this governs the total volume of data that
is being worked on at any given time.

%-----------------------------------------------------------------------
\section{Conclusions}\label{sec:summary}
%-----------------------------------------------------------------------

Modern cosmology experiments have become sufficiently precise and complex that new methods
are required to perform accurate feasibility studies and to perform survey optimization.  We
have described the \spokes\ simulation facility, which is designed to meet the requirements
for future cosmology spectroscopic surveys. The end-to-end architecture of \spokes\ is both
integrated and flexible, and it allows for the reproducibility and modularity needed to
develop, validate and exploit future experiments.

We have demonstrated the completeness, speed and flexibility of the \spokes\ simulation
pipeline. While this was done using the DESpec experiment concept as the baseline for this
purpose, the pipeline is fully general and can be applied to any other wide-field
spectroscopic cosmological experiments. We showed that the pipeline results are consistent
with earlier calculations of the forecast for the science performance of DESpec. We showed
how the \spokes\ framework provides a general tool for detailed studies of systematics,
performance evaluations and development of the data processing pipeline.

In the future, we plan to further develop \spokes. In particular, we plan to implement input
parameters for spectroscopic experiments other than DESpec, to further optimize the modules
to increase performance and parallelisation scaling, and to refine some of the modules with
more physics and more advanced data analysis schemes. In addition, we intend to make
\spokes\ publicly available.
%}{2L}

%-----------------------------------------------------------------------
% Acknowledgments
%-----------------------------------------------------------------------
\section*{Acknowledgments}

The authors would like to thank Brenna Flaugher, Tom Diehl, Michael Meyer, Stephen Bailey,
Richard Kron and Simon Lilly for useful discussions during the development of this work.

%-----------------------------------------------------------------------
% BIBLIOGRAPHY
%-----------------------------------------------------------------------
\bibliography{spokes_pipeline}
\bibliographystyle{elsarticle-harv}

%-----------------------------------------------------------------------
%-----------------------------------------------------------------------
% START APPENDIX
%-----------------------------------------------------------------------
\appendix
%-----------------------------------------------------------------------
%-----------------------------------------------------------------------
%-----------------------------------------------------------------------

\section{Simulation Data}\label{app:sec:simulation}

The galaxy distribution for this mock catalog was created by the AddGals Algorithm. It uses
an input luminosity function to generate a list of galaxies, and then adding the galaxies to
the dark matter simulation using an empirically measured relationship between a galaxies
magnitude, redshift, and local dark matter density, $P(\delta_{dm}|M_r,z)$ -- the
probability that a galaxy with magnitude $M_r$ and redshift $z$ resides in a region with
local density $\delta_{dm}$. This relation was tuned using a high-resolution simulation
combined with the SubHalo Abundance Matching technique that has been shown to reproduce the
observed galaxy 2-point function to high accuracy
\citep{Conroy:2007jp,2013ApJ...771...30R,kravtsov:2008ur}.

For the galaxy assignment algorithm, we choose a luminosity function that is similar to the
SDSS luminosity function as measured in \citep{2003AJ....125.2276B}, but evolves in such a
way as to reproduce the higher redshift observations (e.g., SDSS-Stripe 82, AGES, GAMA,
NDWFS and DEEP2).  In particular, $\phi_*$ and $M_∗$ are varied as a function of redshift in
accordance with the recent results from GAMA \citep{2012MNRAS.420.1239L}.

Once the galaxy positions have been assigned, photometric properties are added. Here, we use
a training set of spectroscopic galaxies taken from SDSS DR5. For each galaxy, in both the
training set and simulation, we measure $\Delta_5$, the distance to the fifth nearest galaxy
on the sky in a redshift bin. Each simulated galaxy is then assigned an SED based on drawing
a random training-set galaxy with the appropriate magnitude and local density, k-correcting
to the appropriate redshift, and projecting onto the desired filters. When doing the color
assignment, the likelihood of assigning a red or a blue galaxy is smoothly varied as a
function of redshift in order simultaneously reproduce the observed red fraction at low and
high redshifts as observed in SDSS and DEEP2.

\section{Data structure}\label{sec:fits_hdf5}

We investigated several data formats to find that which works best for \spokes.

The FITS data format is the most commonly used format for astronomical imaging and catalogs
in the modern era (second perhaps only to ASCII), and has a long history
\citep[e.g.][]{1991BAAS...23..907W}.  We evaluated the FITS format and found that it was not
flexible enough for the requirements of \spokes.  The principal drawback of FITS (and ASCII)
format files is that most i/o functions require reading the entire file before being able to
select particular data fields of interest. This presents large time and memory sinks.  In
addition, most FITS readers do not offer simple access to fields by name (except in
PyFits\footnote{\url{https://pythonhosted.org/pyfits/}}). Partitioning of the data must be
done by an `extension` number (or name), and there is no hierarchical organization
capability.  Finally, there are a finite number of extensions available to have in a FITS
file, and they are organized flatly, not in a nested fashion.

A relational database is very good for a workload that includes mostly read operations, few
write operations and (complex) queries. Sequential processing of all the records in a
database is not optimally performed with relational databases. A relational database is also
an inflexible way to store data, because schema changes are tedious. These features would be
useful for a pipeline with a fixed data set, but not for a pipeline, such as \spokes, that
is built to explore and experiment with different functions and data sets. \spokes\ doesn't
require queries, which would not increase computational efficiency, and it performs few
write operations. \spokes\ also engages primarily in sequential processing of data, which is
not a strength of a relational database.

The HDF5 format permits, by its very nature, hierarchical or nested organization of data via
a unique path, similar to hard disk filesystems. The data sets can be of a variety of data
types, including arrays. According to the HDF Group\footnote{\tt
\url{http://www.hdfgroup.org/why_hdf/}}, HDF supports $n$-dimensional datasets, where any
element can be a complex object. Therefore, we've chosen HDF5 as the data format for the
\spokes\ pipeline.

%-----------------------------------------------------------------------
%-----------------------------------------------------------------------
% END APPENDIX
%-----------------------------------------------------------------------
%-----------------------------------------------------------------------

\end{document}